\documentclass[12pt,a4paper,dvips]{article}
\usepackage{a4p}
\usepackage{cite,mcite}
\usepackage{graphicx}
\usepackage{epsfig}   
\usepackage{physics}
\usepackage{l3_title,ifthen}
\journalname{Phys. Lett. B}

\preprint{2000-118}
\date{August 28, 2000}

\newlength{\capindent}
\newlength{\capwidth}
\newlength{\figwidth}
\newcommand{\icaption}[2][!*!,!]{\hspace*{\capindent}%
  \begin{minipage}{\capwidth}
    \ifthenelse{\equal{#1}{!*!,!}}%
      {\caption{#2}}%
      {\caption[#1]{#2}}
  \end{minipage}}
\newcommand{\phoo}{\phantom{.0}}

\def\tntn{\mathrm{\tau^+\nu_\tau\tau^-\nbar_\tau}}
\def\cstn{\mathrm{c\bar{s} \tau^-\nbar_\tau}}
\def\cscs{\mathrm{c\bar{s} \bar{c}s}}

\def\HHtntn{\mathrm{\Hp\Hm \ra \tntn}}
\def\HHcstn{\mathrm{\Hp\Hm \ra \cstn}}
\def\HHcscs{\mathrm{\Hp\Hm \ra \cscs}}
\def\l{\ifmath{\mathrm{\ell}}}

\def\Htn{\mathrm{H^\pm\ra\tau\nu}}
\def\BRTN{\mathrm{Br(H^\pm\ra\tau\nu)}}
\def\MHPM{\mathrm{m_\Hpm}}

\begin{document}

\begin{titlepage}

\mathversion{bold}
\title{Search for Charged Higgs Bosons in $\bf\ee$ Collisions
       at Centre-of-Mass Energies up to 202 \GeV{}}
\mathversion{normal}

\author{The L3 Collaboration}

\begin{abstract}
  
  A search for pair-produced  charged Higgs bosons is performed with the
  L3 detector at LEP using data  collected  at  centre-of-mass  energies
  between 192 and 202~\GeV{},  corresponding to an integrated luminosity
  of  233.2~\pb.  Decays into a charm and a strange  quark or into a tau
  lepton  and its  neutrino  are  considered.  The  observed  events are
  consistent  with  the  expectations  from  Standard  Model  background
  processes.  Including  data  taken at lower  centre-of-mass  energies,
  lower  limits  on the  charged  Higgs  mass are  derived  at the  95\%
  confidence level.  They vary from 67.4 to 79.9~\GeV{} as a function of
  the $\Htn$ branching ratio.

\end{abstract}

\submitted

\end{titlepage}

%%%%%%%%%%%%%%%%%%%%%%%%%%%%%%%%%%%%%%%%%%%%%%%%%%%%%%%%%%%%%%%%%%%%%%%%%%%%%%%
% Introduction
%%%%%%%%%%%%%%%%%%%%%%%%%%%%%%%%%%%%%%%%%%%%%%%%%%%%%%%%%%%%%%%%%%%%%%%%%%%%%%%

\section*{Introduction}

In    the     Standard     Model~\cite{standard_model},     the    Higgs
mechanism~\cite{higgs_mech}  requires  one  doublet  of  complex  scalar
fields which leads to the  prediction of a single  neutral  scalar Higgs
boson.  Extensions to the minimal  Standard  Model contain more than one
Higgs   doublet~\cite{higgs_hunter}.  In  particular,  models  with  two
complex Higgs doublets  predict two charged Higgs bosons  ($\Hpm$).

A search for the process  $\mathrm{\ee\ra\Hp\Hm}$  is  performed  in the
three      decay      channels       $\mathrm{\Hp\Hm\ra}$       $\tntn$,
$\HHcstn$\footnote{The  charge conjugate  reaction is implied throughout
this letter.}  and  $\HHcscs$,  assumed to be the only possible  decays.
This allows the  interpretation  of the results to be independent of the
$\Htn$ branching ratio.

The  results in this letter are based on data  collected  at  $\sqrt{s}$
between   191.6  and   201.7~\GeV{},   as  well  as  those   from  lower
centre-of-mass  energies, and supersede the previous  lower limit on the
mass    of    the     charged     Higgs     boson     established     by
L3~\cite{l3_48_50,chhiggs_130_183,chhiggs_189}.  Results  from other LEP
experiments   at   lower   centre-of-mass    energies   are   given   in
Reference~\citen{other_lep}.

%%%%%%%%%%%%%%%%%%%%%%%%%%%%%%%%%%%%%%%%%%%%%%%%%%%%%%%%%%%%%%%%%%%%%%%%%%%%%%%
% Data Analysis
%%%%%%%%%%%%%%%%%%%%%%%%%%%%%%%%%%%%%%%%%%%%%%%%%%%%%%%%%%%%%%%%%%%%%%%%%%%%%%%

\section*{Data Analysis}

The search for pair-produced charged Higgs bosons is performed using the
data  collected  in  1999  with  the L3  detector~\cite{l3_det}  at LEP,
corresponding  to  an  integrated   luminosity  of  233.2~\pb{};   where
29.7~\pb{} were collected at a  centre-of-mass  energy of  191.6~\GeV{},
83.7~\pb{} at 195.5~\GeV{}, 82.8~\pb{} at 199.5~\GeV{} and 37.0~\pb{} at
201.7~\GeV{}.  The analyses  remain almost  unchanged since our previous
publications    at    centre-of-mass    energies    between    130   and
189~\GeV~\cite{chhiggs_189,chhiggs_130_183},  with the  exception of the
$\cscs$ final state which is described in more detail below.

The charged Higgs cross section is calculated using the HZHA Monte Carlo
program~\cite{hzha}.  For   the   efficiency   estimates,   samples   of
$\mathrm{\ee\ra\Hp\Hm}$ events are generated with the PYTHIA Monte Carlo
program~\cite{jetset73}  for Higgs  masses  between 50 and  95~\GeV{} in
mass  steps of  5~\GeV{}.  About 1000  events  for each final  state are
generated at each Higgs mass.  For the background studies, the following
Monte  Carlo  generators  are used:  PYTHIA for  $\ee\ra\qqbar(\gamma)$,
$\ee\ra\Zo\Zo$    and    $\ee\ra\Zo\ee$,     KORALW~\cite{KORALW}    for
$\ee\ra\Wp\Wm$,      PHOJET~\cite{PHOJET}     for     $\ee\ra\ee\qqbar$,
DIAG36~\cite{DIAG36}  for  $\ee\ra\ee\ell^+\ell^-$   ($\l=\e,\mu,\tau$),
KORALZ~\cite{KORALZ40}  for $\ee\ra\mu^+\mu^-$ and  $\ee\ra\tau^+\tau^-$
and  BHWIDE~\cite{BHWIDE}  for $\ee\ra\ee$.  The L3 detector response is
simulated  using the  GEANT  program~\cite{my_geant}  which  takes  into
account the effects of energy loss, multiple scattering and showering in
the  detector.  Time-dependent  detector  inefficiencies  are taken into
account in the simulation procedure.

As the theory does not predict  the  branching  ratio for $\Htn$, in the
following  the  performance  of each search  channel is compared  with a
signal  expectation  for a value of $\BRTN$ which is most favourable for
the   corresponding   channel.  This   performance  is  expected  to  be
independent of the quark flavours in the hadronic decay.

\subsection*{{\boldmath{Search in the $\HHtntn$ channel}}}

\begin{figure}[hp]
\centerline{\epsfig{figure=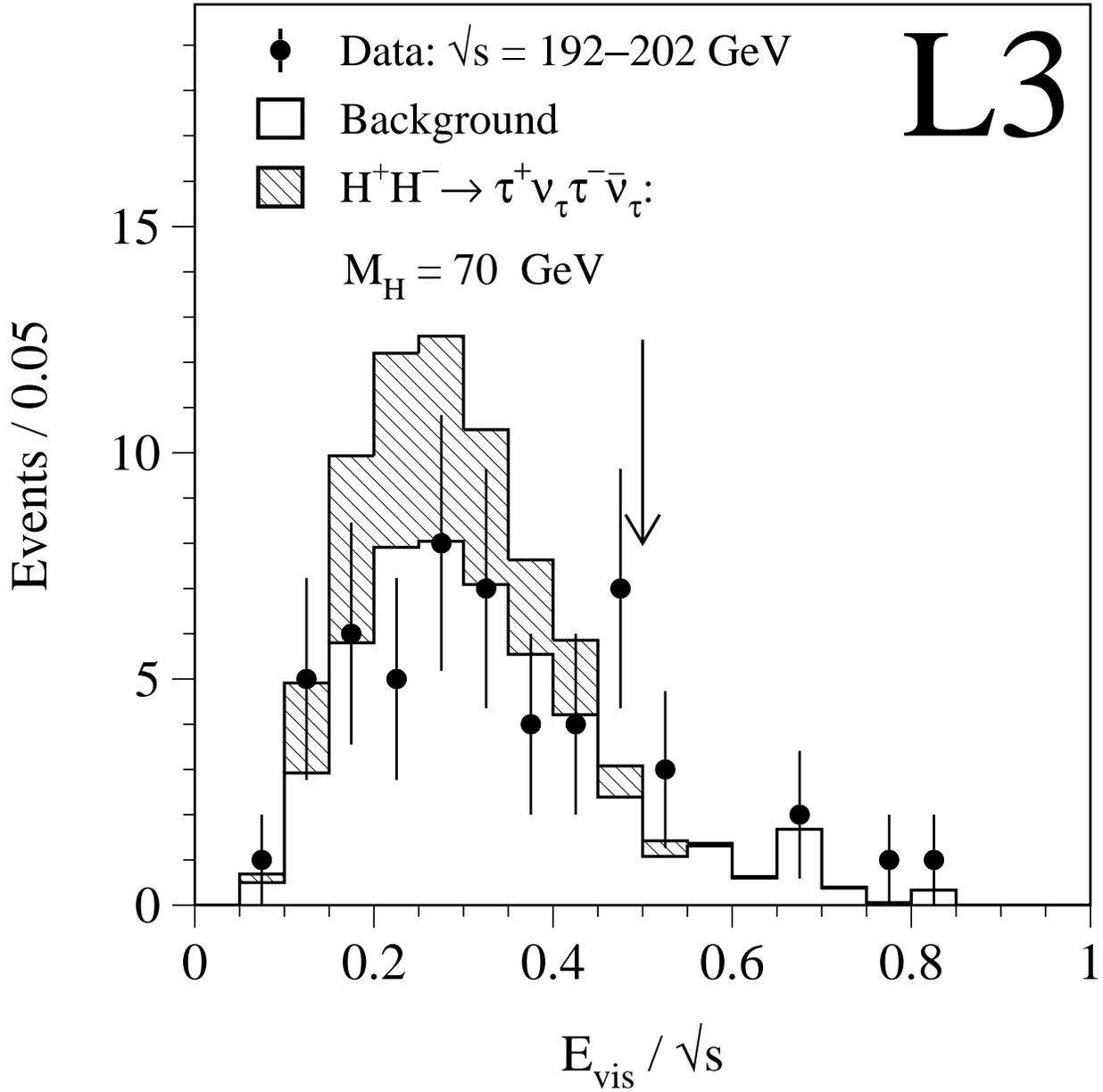,width=1.0\textwidth}}
\caption[]{\label{fig:lepton}  Distribution  of the  normalised  visible
  energy,  $\mathrm{E_{vis}}/\sqrt{s}$,  for the $\HHtntn$ channel after
  all other cuts are  applied.  The arrow  shows the cut  position.  The
  hatched histogram indicates the expected  distribution for a 70~\GeV{}
  Higgs with $\BRTN = 1$.}
\end{figure}

The  signature  for the leptonic  decay channel is a pair of tau leptons
with large missing energy and momentum, giving rise to low  multiplicity
events with low visible energy and a flat distribution in acollinearity,
defined  as  the  maximum   angle   between  any  pair  of  tracks.  The
performance  of the  analysis~\cite{chhiggs_189,chhiggs_130_183}  is not
affected by the increased centre-of-mass energy, and the event selection
remains  unchanged.  Figure~\ref{fig:lepton}  shows the  distribution of
the visible energy for events on which all other selection  criteria are
applied.

The  efficiency of the  $\HHtntn$  selection for several Higgs masses is
listed in Table~\ref{eff}.  The number of data events and the background
expectations  are  presented  in  Table~\ref{events}  for the  different
centre-of-mass  energies.  Almost all the  background  comes from W-pair
production.  The number of events expected for a 70~\GeV{}  Higgs signal
is 19.4 for $\BRTN = 1$.

\subsection*{{\boldmath{Search in the $\HHcstn$ channel}}}

The semileptonic  final state $\HHcstn$ is characterised by two hadronic
jets, a tau lepton and missing momentum.  The selection criteria are the
same   as   for    the    analysis    performed    at    $\sqrt{s}    =$
189~\GeV{}~\cite{chhiggs_189}.

\begin{figure}[hp]
\centerline{\epsfig{figure=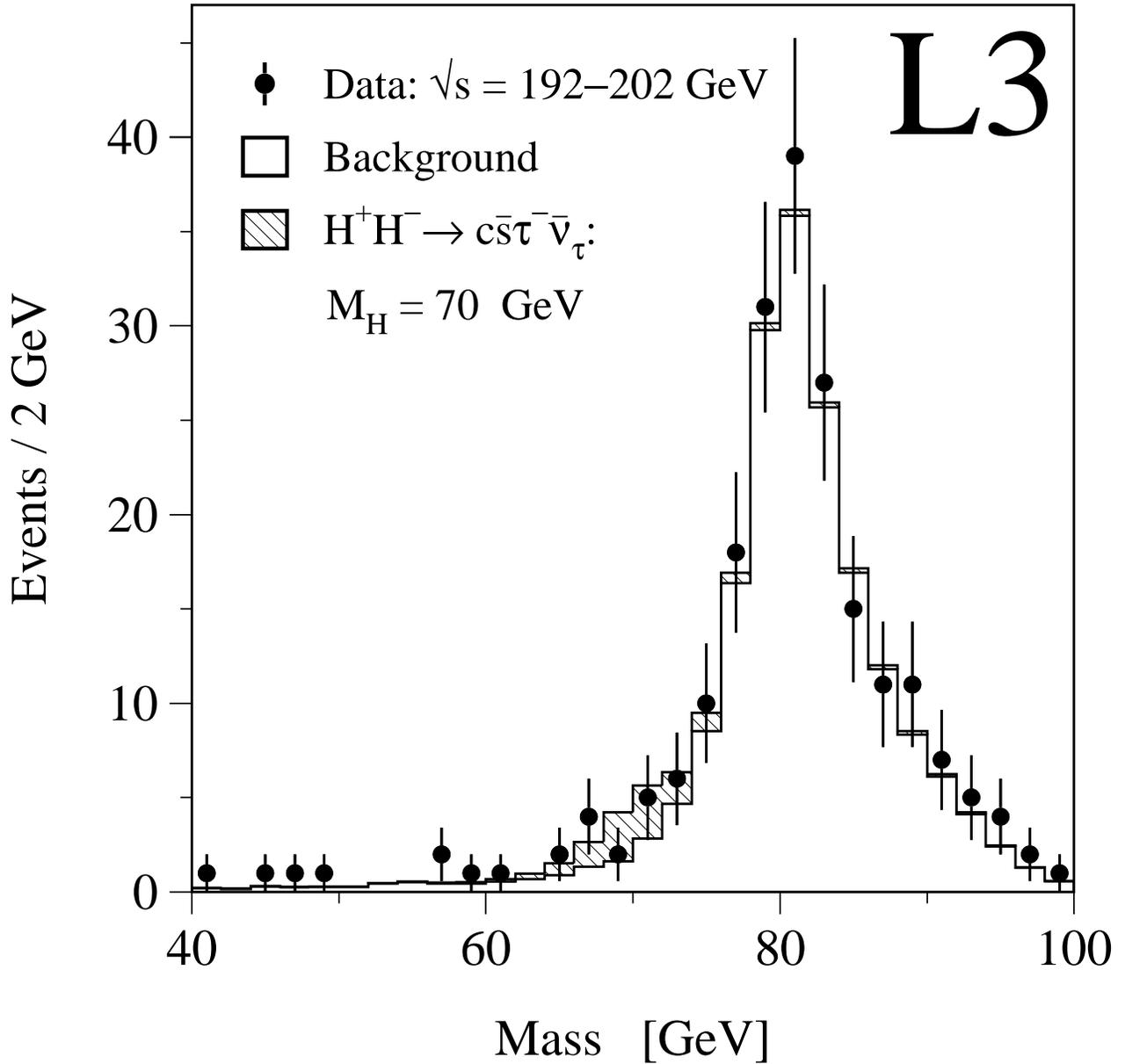,width=1.0\textwidth}}
\caption[]{\label{fig:cstn}  Reconstructed  mass  spectrum  for data and
  expected   background   in  the   $\HHcstn$   channel.  The   expected
  distribution  for a 70~\GeV{} Higgs with $\BRTN = 0.5$ is added as the
  hatched histogram.}
\end{figure}

The selection  efficiencies are given in  Table~{\ref{eff}}.  The number
of data events and the background expected from Standard Model processes
are  listed  in  Table~\ref{events}  for  the  different  centre-of-mass
energies.   The    background    is    dominated    by    the    process
$\Wp\Wm\ra\qqbar^\prime\tau\nu$.  The  number of events  expected  for a
70~\GeV{} Higgs signal is 13.2 for $\BRTN = 0.5$.  Figure~\ref{fig:cstn}
displays the distribution of the average of the jet-jet and $\tau$-$\nu$
masses.  They are  calculated  from a kinematic fit imposing  energy and
momentum conservation for an assumed production of equal mass particles,
keeping the  directions  of the jets, the tau and the  missing  momentum
vector at their measured values.

\begin{table}
\begin{center}
\begin{tabular}{|l|ccccc|r|r|} \hline
\multicolumn{1}{|c|}{\raisebox{-8pt}[0pt][0pt]{Channel}}
         & \multicolumn{5}{c|}{Selection efficiency (\%) for $ \MHPM = $} \\
         & 60 \GeV  & 65 \GeV & 70 \GeV & 75 \GeV  & 80 \GeV    \\ \hline
$\tntn$  &   28     &   30    &  31     &   32     &   33       \\
$\cstn$  &   42     &   42    &  43     &   41     &   37       \\
$\cscs$  &   51     &   56    &  60     &   61     &   63       \\ \hline
\end{tabular}
\caption{The  charged Higgs  selection  efficiencies  for various  Higgs
    masses, averaged over centre-of-mass  energies and weighted with the
    luminosity.  The   efficiencies   are  almost   independent  of  the
    centre-of-mass   energy.  The  uncertainty  on  each  efficiency  is
    estimated to be 3\%.}
\label{eff}
\end{center}
\end{table}

\begin{table}
\begin{center}
\begin{tabular}{|l|rrrr|c|l|} \cline{1-6}
\multicolumn{1}{|c|}{\raisebox{-8pt}[0pt][0pt]{Channel}}
         & \multicolumn{4}{c|}{Centre-of-mass energy (\GeV{})}       &\quad \raisebox{-8pt}[0pt][0pt]{Total} \quad& \multicolumn{1}{|c}{} \\
         &\quad 191.6   &\quad 195.5   &\quad 199.5   &\quad 201.7   &                 & \multicolumn{1}{|c}{} \\ \hline
$\tntn$  &    6\phoo    &   23\phoo    &   12\phoo    &    6\phoo    &     47\phoo     & Data                  \\
         &    5.4       &   15.4       &   16.3       &    6.9       &     44.0        & Background            \\
         &              &              &              &              &                 &                       \\
$\cstn$  &   21\phoo    &   88\phoo    &   66\phoo    &   34\phoo    &    209\phoo     & Data                  \\
         &   22.8       &   67.2       &   64.0       &   30.4       &    184.4        & Background            \\
         &              &              &              &              &                 &                       \\
$\cscs$  &  133\phoo    &  379\phoo    &  384\phoo    &  146\phoo    &   1042\phoo     & Data                  \\
         &  139.1       &  384.9       &  364.6       &  162.5       &   1051.1        & Background            \\ \hline
\end{tabular}
\caption{The  number of data events and the background  expectations per
    centre-of-mass   energy.   The   uncertainty   on   the   background
    expectations is estimated to be 6\%.}
\label{events}
\end{center}
\end{table}

\subsection*{{\boldmath{Search in the $ \HHcscs$ channel}}}

Events  from the  $\HHcscs$  channel  have a high  multiplicity  and are
balanced in transverse  and  longitudinal  momenta.  A large fraction of
the  centre-of-mass  energy is deposited in the  detector,  typically as
four hadronic jets.  The selection  criteria are slightly  modified with
respect     to     the     analysis     at     lower      centre-of-mass
energies~\cite{chhiggs_189,chhiggs_130_183},    in    order    to   gain
sensitivity at higher masses.

Events with an identified electron, muon or photon with energy in excess
of 65~\GeV{} are discarded.  A neural  network~\cite{nnet} is applied to
distinguish  events  with four  genuine  quark jets from those  with two
quark jets and two jets from gluon radiation.  Further  reduction of the
QCD  background  is  achieved  by  requiring  the Durham jet  resolution
parameter,  $y_{34}$,  for which  three-jet  events  are  resolved  into
four-jet ones, to be greater than 0.003 and by requiring the minimum jet
energy to exceed 6\% of $\sqrt{s}$.

The charged Higgs  production angle  distribution  has a  $\sin^2\theta$
dependence,  where  $\theta$ is the polar angle with respect to the beam
direction.  A cut of $|  \cos\theta  | < 0.8 $ is  therefore  applied to
preferentially reject W-pair background.

The analysis of the  55.3~\pb{}  and  176.4~\pb{}  of data  respectively
taken at $\sqrt{s}  =$ 183 and 189 \GeV{} is redone  using the  criteria
described         above,         superseding         the        previous
analysis~\cite{chhiggs_189,chhiggs_130_183}.  The   number   of   events
selected in data at these centre-of-mass  energies is 1103, while 1085.7
background events are expected from Standard Model processes.

\begin{figure}[hp]
\centerline{\epsfig{figure=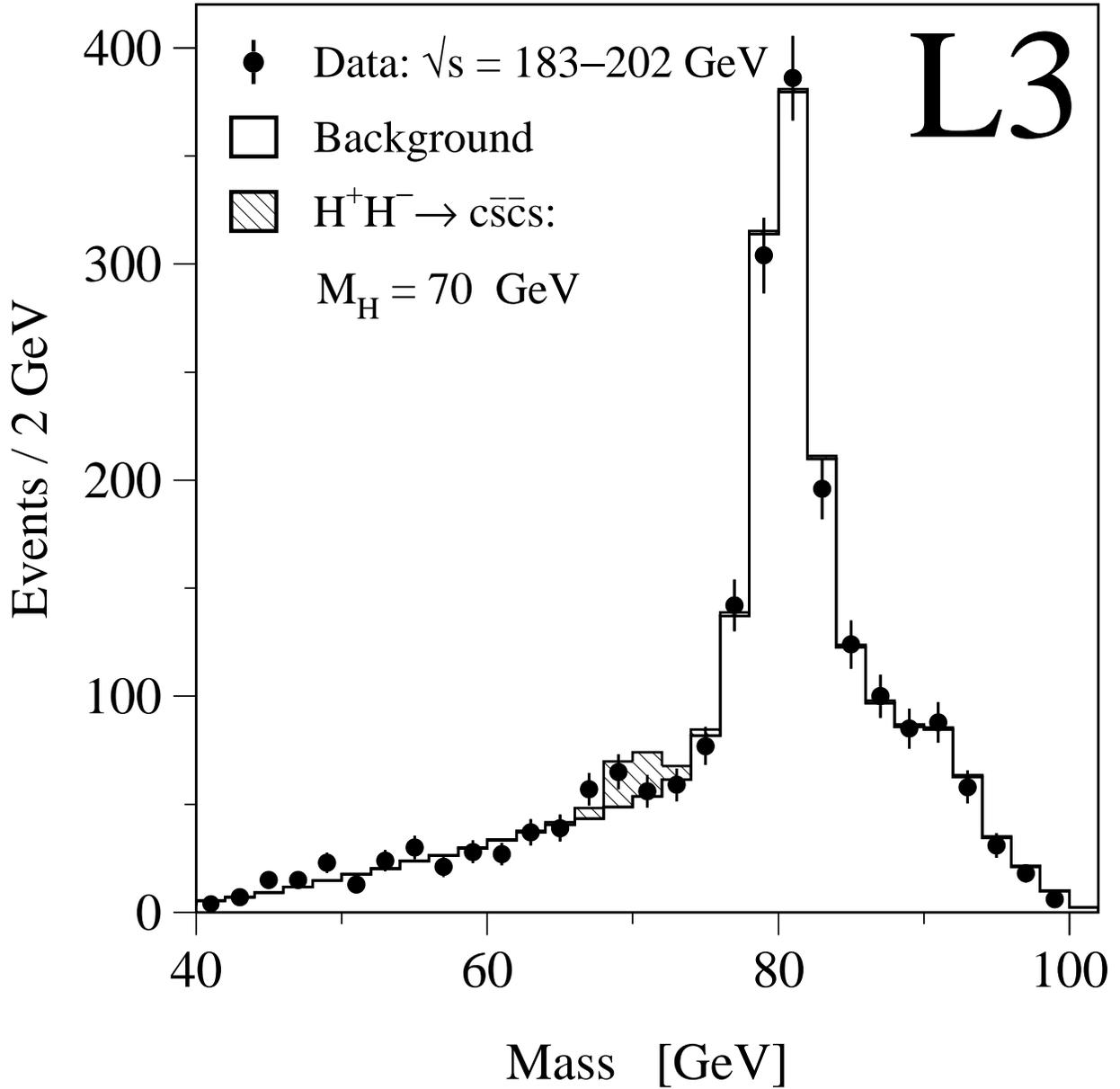,width=1.0\textwidth}}
\caption[]{\label{fig:cscs}  Distribution  of the mass  resulting from a
  kinematic fit, assuming  production of equal mass  particles, for data
  and  expected   background  in  the  $\HHcscs$  channel.  The  hatched
  histogram  indicates the expected  distribution  for a 70~\GeV{} Higgs
  with $\BRTN = 0$.}
\end{figure}

The selection efficiencies are listed in Table~\ref{eff}.  The number of
events  selected in data and the  background  expectations  are given in
Table~\ref{events} for the different  centre-of-mass energies.  The main
contribution to the background  comes from W-pair decays into four jets.
The number of events  expected for a 70~\GeV{}  Higgs signal is 37.4 for
$\BRTN = 0$.  Figure~\ref{fig:cscs}  shows the dijet  mass  distribution
after a kinematic  fit  imposing  four-momentum  conservation  and equal
dijet masses.  A slight excess is observed around 68~\GeV{}.

%%%%%%%%%%%%%%%%%%%%%%%%%%%%%%%%%%%%%%%%%%%%%%%%%%%%%%%%%%%%%%%%%%%%%%%%%%%%%%%
% Conclusion
%%%%%%%%%%%%%%%%%%%%%%%%%%%%%%%%%%%%%%%%%%%%%%%%%%%%%%%%%%%%%%%%%%%%%%%%%%%%%%%

\section*{Results}

\begin{figure}[hp]
\centerline{\epsfig{figure=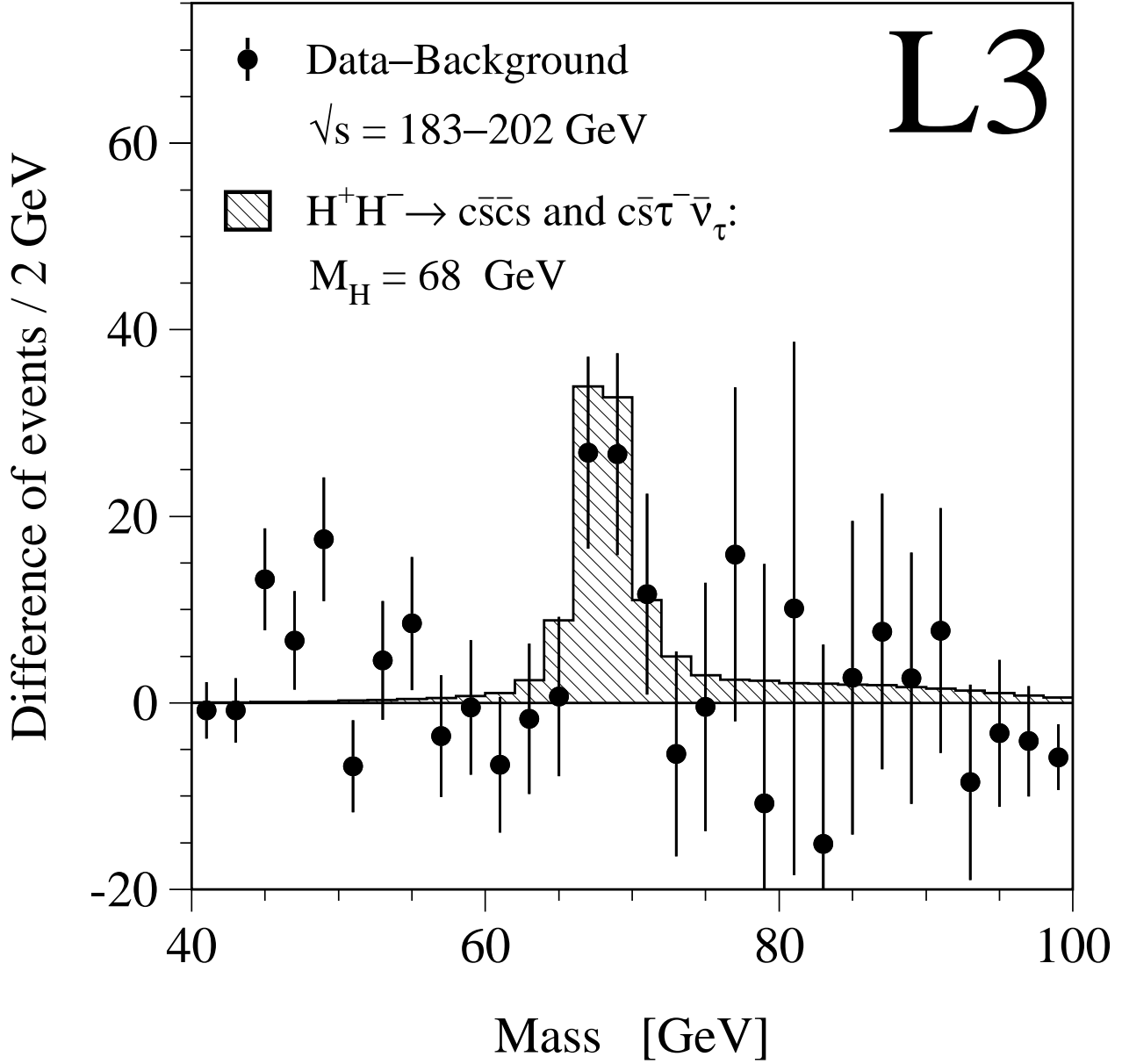,width=1.0\textwidth}}
\caption[]{\label{fig:mass}    Combined    background-subtracted    mass
  distribution  for the $\HHcscs$ and $\cstn$ decay channels,  where the
  events are corrected for the efficiency of their respective  analyses.
  The expected  distribution for a 68~\GeV{} Higgs with $\BRTN = 0.1$ is
  shown by the hatched histogram.}
\end{figure}

The number of selected  events in each decay channel is consistent  with
the number of events  expected from Standard Model  processes.  However,
there  is  an  excess  of  events  in  the  $\cscs$  and  $\cstn$   mass
distributions  around  68~\GeV{}.  Figure~\ref{fig:mass}   displays  the
combined  background-subtracted  mass  distribution  for these two Higgs
decay  channels,  where the events are corrected  for the  efficiency of
their   respective   analyses.  The  figure  also  shows  the   expected
distribution  for a  68~\GeV{}  Higgs with $\BRTN = 0.1$.  This value of
the  $\BRTN$  is in the  range of  branching  fractions  for  which  the
observed  excess of events  is  closest  to the  expected  number  for a
68~\GeV{} mass Higgs, as is described in more detail below.

In order to estimate the  significance  of this  excess, the  systematic
errors on the background and signal efficiencies are taken into account.
The main  systematic  uncertainties  come from the  finite  Monte  Carlo
statistics  and the precision of the cross  sections for the  background
processes.  The former is calculated for each final state using binomial
statistics,  leading to an overall 5\% uncertainty in the background and
3\% in the signal  normalisation.  The latter  affects the  analyses  in
different   ways   depending   on  the   background   composition.  This
uncertainty  is 2\% for the $\HHtntn$ and $\cstn$  channels, and 3\% for
$\cscs$.  The systematic uncertainty on the signal efficiency due to the
selection  is  estimated to be less than 1\%, by varying the cut values.
The total  systematic  error on the number of  expected  background  and
signal events is therefore estimated to be 6\% and 3\%, respectively.

A  technique  based  on a  log-likelihood  ratio~\cite{lnq}  is  used to
calculate  a  confidence   level  (CL)  that  the  observed  events  are
consistent with background  expectations.  The  test-statistic  adopted,
$Q$, is the  ratio  of the  likelihood  function  for  the  signal  plus
background hypothesis to the likelihood function for the background only
hypothesis.  For the $\cscs$ and  $\cstn$  channels,  the  reconstructed
mass distributions (Figures~\ref{fig:cstn}  and~\ref{fig:cscs}) are used
in the calculation, whereas for the $\tntn$ channel, the total number of
data,  expected  background  and expected  signal  events are used.  The
systematic  uncertainties on the background and signal  efficiencies are
included in the confidence level calculation.

\begin{figure}[hp]
\centerline{\epsfig{figure=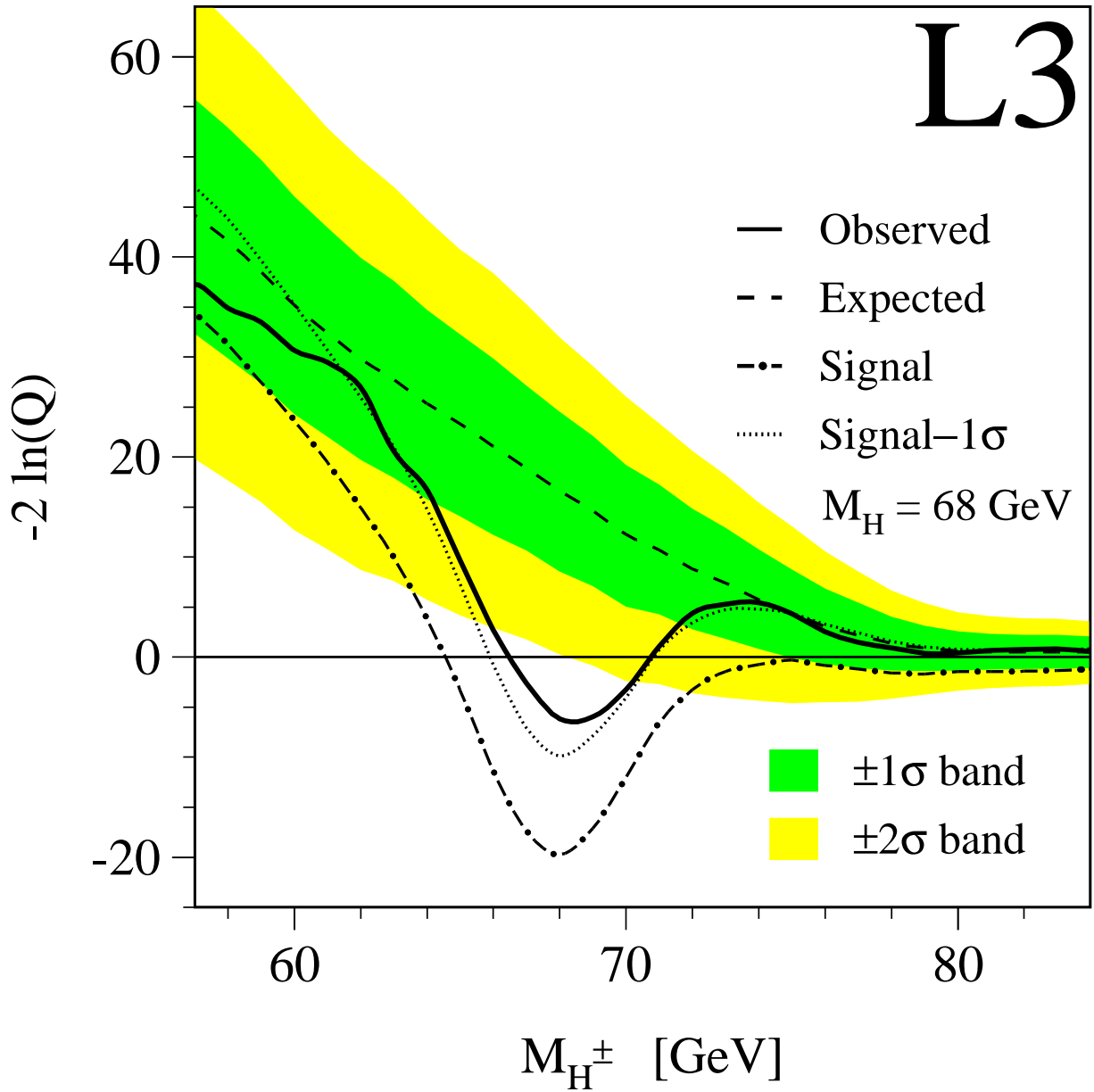,width=1.0\textwidth}}
\caption[]{\label{fig:lnq}   The  negative   log-likelihood  ratio,  $-2
  \ln{(Q)}$,  as a function  of the Higgs mass with  $\BRTN = 0.1$.  The
  solid line shows the values computed from the observed results and the
  dashed line the expectation for the  background only  hypothesis.  The
  dash-dotted line is the curve expected for a 68 \GeV{} Higgs signal at
  this value of the  branching  ratio.  The dotted line is the  expected
  result for a $1 \sigma$  under-fluctuation  of the  signal. The shaded
  areas  represent the  symmetric $1 \sigma$ and $2 \sigma$  probability
  bands expected in the absence of a signal.}
\end{figure}

Figure~\ref{fig:lnq}  shows the resulting negative log-likelihood ratio,
$-2 \ln{(Q)}$, using $\BRTN = 0.1$, as a function of the Higgs mass, for
the data and for the  expectation  in the  absence of a signal.  The one
and two standard deviation ($\sigma$)  probability bands expected in the
absence of a signal  are also  displayed.  The  excess of events  around
$\MHPM = 68$~\GeV{} is compatible with a $2.7\sigma$  fluctuation in the
background.  The  statistical  significance  of  the  excess  is  almost
constant    for    values   of    $\BRTN$    between    0.1   and   0.2.
Figure~\ref{fig:lnq}  also shows the expected $-2 \ln{(Q)}$ distribution
for the  hypothesis of a 68~\GeV{}  mass Higgs signal and its  $1\sigma$
under-fluctuation.  The data are $1.4\sigma$  below what is expected for
a Higgs  signal at this mass.  Again, this  difference  is not  strongly
dependent on the value of the branching fraction.

\begin{figure}[hp]
\centerline{\epsfig{figure=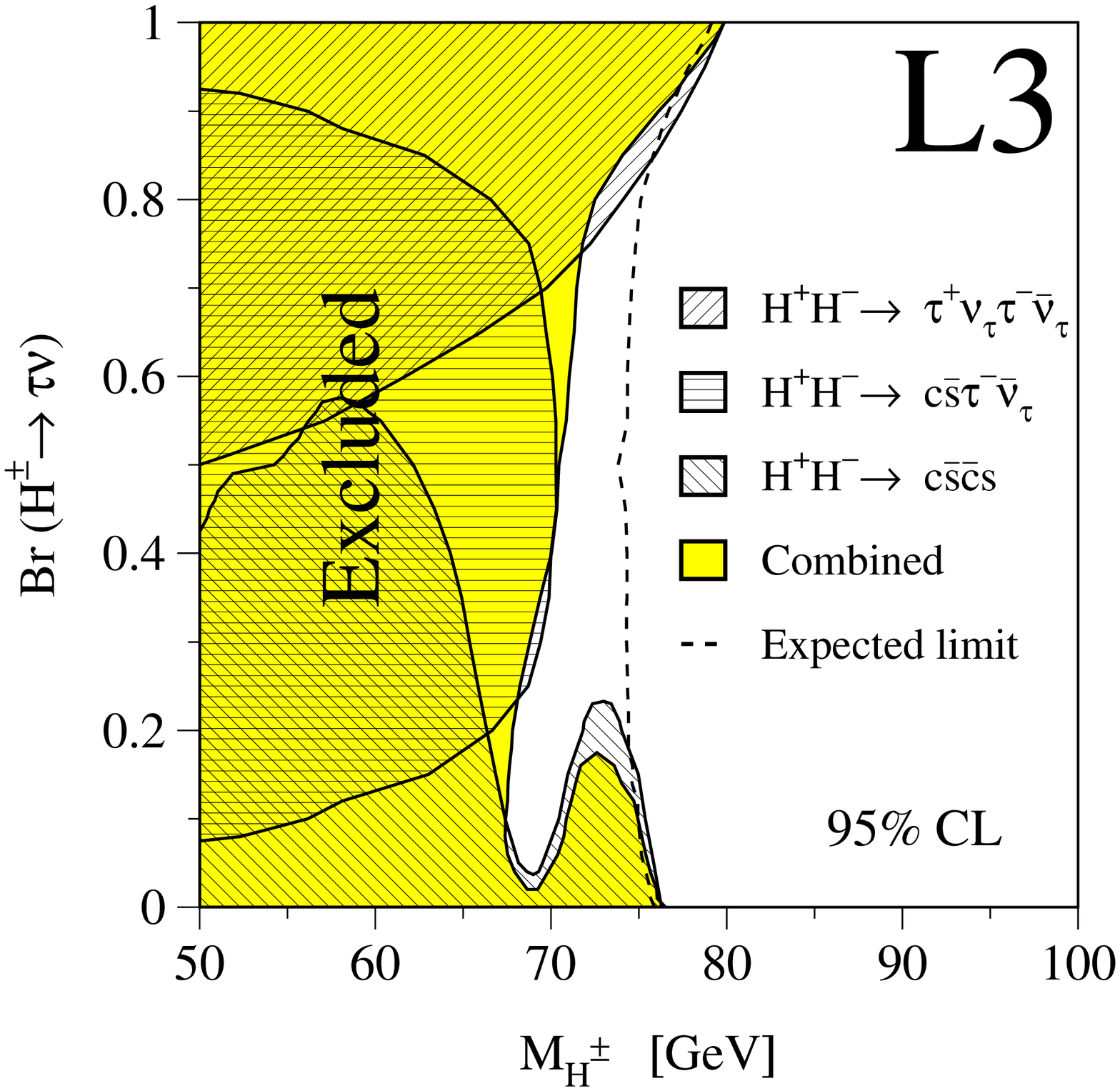,width=1.0\textwidth}}
\caption[]{\label{exclusion}  Excluded  regions  for the  charged  Higgs
  boson  at more  than  95\% CL in the  plane  of the  $\Htn$  branching
  fraction  versus  mass, for the analyses of each final state and their
  combination.  The dashed line indicates the median  expected  limit in
  the absence of a signal.  There are regions excluded by the individual
  analyses but not by their combination.}
\end{figure}

Interpreting this excess as a statistical fluctuation in the background,
lower limits on the charged  Higgs mass as a function of the $\BRTN$ are
derived~\cite{lnq,l3cl}  at the 95\% CL, using the data from  $\sqrt{s}$
between   191.6  and   201.7~\GeV{},   as  well  as  those   from  lower
centre-of-mass              energies~\cite{chhiggs_189,chhiggs_130_183}.
Figure~\ref{exclusion} shows the excluded Higgs mass regions for each of
the final states and their  combination,  as a function of the  $\BRTN$.
Some regions which are excluded  using one channel are not excluded when
all three channels are combined.  Table~\ref{cl}  gives the observed and
the median  expected  lower limits for several  values of the  branching
ratio.  The  region  around  $\MHPM =  68$~\GeV{}  at low  values of the
$\BRTN$  can  only be  excluded  at 88\%  CL, due to the  aforementioned
excess of events in this mass  region.  A similar  but less  significant
excess was observed in our previous publication~\cite{chhiggs_189}.

\begin{table}
\begin{center}
\begin{tabular}{|c|cc|} \hline
\raisebox{-8pt}[0pt][0pt]{$\BRTN$}
       & \multicolumn{2}{|c|}{Lower limits at 95\% CL (\GeV{})} \\
       & \ observed \      & median expected                    \\ \hline
  0.0  &        76.5       &        75.9                        \\
  0.1  &        67.4       &        74.9                        \\
  0.5  &        70.5       &        73.8                        \\
  1.0  &        79.9       &        79.2                        \\ \hline
\end{tabular}
\caption{Observed  and  median  expected  lower  limits  at 95\%  CL for
    different   values  of  the  $\Htn$  branching  ratio.  The  minimum
    observed limit, independent of the branching  fraction, is at $\BRTN
    = 0.1$.}
\label{cl}
\end{center}
\end{table}

Our  sensitivity  to larger Higgs  masses, as  quantified  by the median
expected mass limits given in Table~\ref{cl}, is significantly  improved
as  compared   with  our  previous   results  at  lower   centre-of-mass
energies~\cite{chhiggs_189,chhiggs_130_183}.  Combining all our data, we
obtain  a new  lower  limit  at  95\%  CL of  $$\MHPM  > 67.4  \GeV{},$$
independent of the branching ratio.

\section*{Acknowledgements}

We wish to express our gratitude to the CERN  accelerator  divisions for
the  excellent  performance  of the  LEP  machine.  We  acknowledge  the
contributions of the engineers and technicians who have  participated in
the construction and maintenance of this experiment.

%%%%%%%%%%%%%%%%%%%%%%%%%%%%%%%%%%%%%%%%%%%%%%%%%%%%%%%%%%%%%%%%%%%%%%%%%%%%%%%
% Bibliography
%%%%%%%%%%%%%%%%%%%%%%%%%%%%%%%%%%%%%%%%%%%%%%%%%%%%%%%%%%%%%%%%%%%%%%%%%%%%%%
% Style file to use with mcite.
% Use l3style with just cite.

%%%%%%%%%%%%%%%%%%%%%%%%%%%%%%%%%%%%%%%%%%%%%%%%%%%%%%%%%%%%%%%%%%%%%%%%%%%%%%%
% The author list
%%%%%%%%%%%%%%%%%%%%%%%%%%%%%%%%%%%%%%%%%%%%%%%%%%%%%%%%%%%%%%%%%%%%%%%%%%%%%%%
%
\newpage

\newcount\tutecount  \tutecount=0
\def\tutenum#1{\global\advance\tutecount by 1 \xdef#1{\the\tutecount}}
\def\tute#1{$^{#1}$}
\tutenum\aachen            % 1
\tutenum\nikhef            % 2
\tutenum\mich              % 3
\tutenum\lapp              % 4
\tutenum\basel             % 5
\tutenum\lsu               % 6
\tutenum\beijing           % 7
\tutenum\berlin            % 8
\tutenum\bologna           % 9 
\tutenum\tata              % 10
\tutenum\ne                % 11
\tutenum\bucharest         % 12
\tutenum\budapest          % 13
\tutenum\mit               % 14 
\tutenum\debrecen          % 15
\tutenum\florence          % 16
\tutenum\cern              % 17 
\tutenum\wl                % 18 
\tutenum\geneva            % 19
\tutenum\hefei             % 20
\tutenum\seft              % 21
\tutenum\lausanne          % 22
\tutenum\lecce             % 23
\tutenum\lyon              % 24
\tutenum\madrid            % 25
\tutenum\milan             % 26
\tutenum\moscow            % 27
\tutenum\naples            % 27
\tutenum\cyprus            % 29
\tutenum\nymegen           % 30
\tutenum\caltech           % 31
\tutenum\perugia           % 32
\tutenum\cmu               % 33
\tutenum\prince            % 34
\tutenum\rome              % 35
\tutenum\peters            % 36
\tutenum\potenza           % 37
\tutenum\salerno           % 38
\tutenum\ucsd              % 39
\tutenum\santiago          % 40
\tutenum\sofia             % 41 
\tutenum\korea             % 42
\tutenum\alabama           % 43
\tutenum\utrecht           % 44
\tutenum\purdue            % 45
\tutenum\psinst            % 46
\tutenum\zeuthen           % 47
\tutenum\eth               % 48
\tutenum\hamburg           % 49
\tutenum\taiwan            % 50
\tutenum\tsinghua          % 51

{
\parskip=0pt
\noindent
{\bf The L3 Collaboration:}
\ifx\selectfont\undefined%  old style font selection
 \baselineskip=10.8pt
 \baselineskip\baselinestretch\baselineskip
 \normalbaselineskip\baselineskip
 \ixpt
\else%                      new style font selection
 \fontsize{9}{10.8pt}\selectfont
\fi
\medskip
\tolerance=10000
\hbadness=5000
\raggedright
\hsize=162truemm\hoffset=0mm
\def\r{\rlap,}
\noindent

M.Acciarri\r\tute\milan\
P.Achard\r\tute\geneva\ 
O.Adriani\r\tute{\florence}\ 
M.Aguilar-Benitez\r\tute\madrid\ 
J.Alcaraz\r\tute\madrid\ 
G.Alemanni\r\tute\lausanne\
J.Allaby\r\tute\cern\
A.Aloisio\r\tute\naples\ 
M.G.Alviggi\r\tute\naples\
G.Ambrosi\r\tute\geneva\
% ...
\newpage
%\rule{\textwidth}{0.4pt}
\begin{list}{A}{\itemsep=0pt plus 0pt minus 0pt\parsep=0pt plus 0pt minus 0pt
                \topsep=0pt plus 0pt minus 0pt}
\item[\aachen]
 I. Physikalisches Institut, RWTH, D-52056 Aachen, FRG$^{\S}$\\
 III. Physikalisches Institut, RWTH, D-52056 Aachen, FRG$^{\S}$
\item[\nikhef] National Institute for High Energy Physics, NIKHEF, 
     and University of Amsterdam, NL-1009 DB Amsterdam, The Netherlands
\item[\mich] University of Michigan, Ann Arbor, MI 48109, USA
\item[\lapp] Laboratoire d'Annecy-le-Vieux de Physique des Particules, 
     LAPP,IN2P3-CNRS, BP 110, F-74941 Annecy-le-Vieux CEDEX, France
\item[\basel] Institute of Physics, University of Basel, CH-4056 Basel,
     Switzerland
\item[\lsu] Louisiana State University, Baton Rouge, LA 70803, USA
\item[\beijing] Institute of High Energy Physics, IHEP, 
  100039 Beijing, China$^{\triangle}$ 
\item[\berlin] Humboldt University, D-10099 Berlin, FRG$^{\S}$
\item[\bologna] University of Bologna and INFN-Sezione di Bologna, 
     I-40126 Bologna, Italy
\item[\tata] Tata Institute of Fundamental Research, Bombay 400 005, India
\item[\ne] Northeastern University, Boston, MA 02115, USA
\item[\bucharest] Institute of Atomic Physics and University of Bucharest,
     R-76900 Bucharest, Romania
\item[\budapest] Central Research Institute for Physics of the 
     Hungarian Academy of Sciences, H-1525 Budapest 114, Hungary$^{\ddag}$
\item[\mit] Massachusetts Institute of Technology, Cambridge, MA 02139, USA
\item[\debrecen] KLTE-ATOMKI, H-4010 Debrecen, Hungary$^\P$
\item[\florence] INFN Sezione di Firenze and University of Florence, 
     I-50125 Florence, Italy
\item[\cern] European Laboratory for Particle Physics, CERN, 
     CH-1211 Geneva 23, Switzerland
\item[\wl] World Laboratory, FBLJA  Project, CH-1211 Geneva 23, Switzerland
\item[\geneva] University of Geneva, CH-1211 Geneva 4, Switzerland
\item[\hefei] Chinese University of Science and Technology, USTC,
      Hefei, Anhui 230 029, China$^{\triangle}$
\item[\seft] SEFT, Research Institute for High Energy Physics, P.O. Box 9,
      SF-00014 Helsinki, Finland
\item[\lausanne] University of Lausanne, CH-1015 Lausanne, Switzerland
\item[\lecce] INFN-Sezione di Lecce and Universit\'a Degli Studi di Lecce,
     I-73100 Lecce, Italy
\item[\lyon] Institut de Physique Nucl\'eaire de Lyon, 
     IN2P3-CNRS,Universit\'e Claude Bernard, 
     F-69622 Villeurbanne, France
\item[\madrid] Centro de Investigaciones Energ{\'e}ticas, 
     Medioambientales y Tecnolog{\'\i}cas, CIEMAT, E-28040 Madrid,
     Spain${\flat}$ 
\item[\milan] INFN-Sezione di Milano, I-20133 Milan, Italy
\item[\moscow] Institute of Theoretical and Experimental Physics, ITEP, 
     Moscow, Russia
\item[\naples] INFN-Sezione di Napoli and University of Naples, 
     I-80125 Naples, Italy
\item[\cyprus] Department of Natural Sciences, University of Cyprus,
     Nicosia, Cyprus
\item[\nymegen] University of Nijmegen and NIKHEF, 
     NL-6525 ED Nijmegen, The Netherlands
\item[\caltech] California Institute of Technology, Pasadena, CA 91125, USA
\item[\perugia] INFN-Sezione di Perugia and Universit\'a Degli 
     Studi di Perugia, I-06100 Perugia, Italy   
\item[\cmu] Carnegie Mellon University, Pittsburgh, PA 15213, USA
\item[\prince] Princeton University, Princeton, NJ 08544, USA
\item[\rome] INFN-Sezione di Roma and University of Rome, ``La Sapienza",
     I-00185 Rome, Italy
\item[\peters] Nuclear Physics Institute, St. Petersburg, Russia
\item[\potenza] INFN-Sezione di Napoli and University of Potenza, 
     I-85100 Potenza, Italy
\item[\salerno] University and INFN, Salerno, I-84100 Salerno, Italy
\item[\ucsd] University of California, San Diego, CA 92093, USA
\item[\santiago] Dept. de Fisica de Particulas Elementales, Univ. de Santiago,
     E-15706 Santiago de Compostela, Spain
\item[\sofia] Bulgarian Academy of Sciences, Central Lab.~of 
     Mechatronics and Instrumentation, BU-1113 Sofia, Bulgaria
\item[\korea]  Laboratory of High Energy Physics, 
     Kyungpook National University, 702-701 Taegu, Republic of Korea
\item[\alabama] University of Alabama, Tuscaloosa, AL 35486, USA
\item[\utrecht] Utrecht University and NIKHEF, NL-3584 CB Utrecht, 
     The Netherlands
\item[\purdue] Purdue University, West Lafayette, IN 47907, USA
\item[\psinst] Paul Scherrer Institut, PSI, CH-5232 Villigen, Switzerland
\item[\zeuthen] DESY, D-15738 Zeuthen, 
     FRG
\item[\eth] Eidgen\"ossische Technische Hochschule, ETH Z\"urich,
     CH-8093 Z\"urich, Switzerland
\item[\hamburg] University of Hamburg, D-22761 Hamburg, FRG
\item[\taiwan] National Central University, Chung-Li, Taiwan, China
\item[\tsinghua] Department of Physics, National Tsing Hua University,
      Taiwan, China
\item[\S]  Supported by the German Bundesministerium 
        f\"ur Bildung, Wissenschaft, Forschung und Technologie
\item[\ddag] Supported by the Hungarian OTKA fund under contract
numbers T019181, F023259 and T024011.
\item[\P] Also supported by the Hungarian OTKA fund under contract
  numbers T22238 and T026178.
\item[$\flat$] Supported also by the Comisi\'on Interministerial de Ciencia y 
        Tecnolog{\'\i}a.
\item[$\sharp$] Also supported by CONICET and Universidad Nacional de La Plata,
        CC 67, 1900 La Plata, Argentina.
\item[$\diamondsuit$] Also supported by Panjab University, Chandigarh-160014, 
        India.
\item[$\triangle$] Supported by the National Natural Science
  Foundation of China.
\end{list}
}
\vfill

%%% Local Variables: 
%%% mode: latex
%%% TeX-master: t
%%% End:

%%%%%%%%%%%%%%%%%%%%%%%%%%%%%%%%%%%%%%%%%%%%%%%%%%%%%%%%%%%%%%%%%%%%%%%%%%%%%%%
%\subsection*{Figures}
%%%%%%%%%%%%%%%%%%%%%%%%%%%%%%%%%%%%%%%%%%%%%%%%%%%%%%%%%%%%%%%%%%%%%%%%%%%%%%%

% \newpage

~

\begin{mcbibliography}{10}

\bibitem{standard_model}
S.L. Glashow, \NP {\bf 22} (1961) 579;\\ S. Weinberg, \PRL {\bf 19} (1967)
  1264;\\ A. Salam, {\em Elementary Particle Theory}, edited by N.~Svartholm
  (Almqvist and Wiksell, Stockholm, 1968), p. 367\relax
\relax
\bibitem{higgs_mech}
P.W. Higgs, \PL {\bf 12} (1964) 132,~\PRL {\bf 13} (1964) 508 and \PR {\bf 145}
  (1966) 1156;\\ F.~Englert and R.~Brout, \PRL {\bf 13} (1964) 321;\\ G.S.
  Guralnik, C.R. Hagen and T.W.B. Kibble, Phys. Rev. Lett. {\bf 13} (1964)
  585\relax
\relax
\bibitem{higgs_hunter}
S. Dawson \etal, {\em The Physics of the Higgs Bosons: Higgs Hunter's Guide},
  Addison Wesley, Menlo Park, 1989\relax
\relax
\bibitem{l3_48_50}
L3 Collab., O.~Adriani \etal, \PL {\bf B 294} (1992) 457; \\ L3 Collab.,
  O.~Adriani \etal, \ZfP {\bf C 57} (1993) 355\relax
\relax
\bibitem{chhiggs_130_183}
L3 Collab., M.~Acciarri \etal, Phys. Lett. {\bf B 446} (1999) 368\relax
\relax
\bibitem{chhiggs_189}
L3 Collab., M.~Acciarri \etal, Phys. Lett. {\bf B 466} (1999) 71\relax
\relax
\bibitem{other_lep}
ALEPH Collab., R. Barate \etal, Phys. Lett. {\bf B 418} (1998) 419;\\ ALEPH
  Collab., R. Barate \etal, Phys. Lett. {\bf B 450} (1999) 467;\\ DELPHI
  Collab., P. Abreu \etal, Phys. Lett. {\bf B 420} (1998) 140;\\ OPAL Collab.,
  K. Ackerstaff \etal, Phys. Lett. {\bf B 426} (1998) 180;\\ OPAL Collab., G.
  Abbiendi \etal, Eur. Phys. J. {\bf C 7} (1999) 407\relax
\relax
\bibitem{l3_det}
L3 Collab., B. Adeva \etal, Nucl. Instr. Meth. {\bf A 289} (1990) 35;\\ J.A.
  Bakken \etal, Nucl. Instr. Meth. {\bf A 275} (1989) 81;\\ O. Adriani \etal,
  Nucl. Instr. Meth. {\bf A 302} (1991) 53;\\ B. Adeva \etal, Nucl. Instr.
  Meth. {\bf A 323} (1992) 109;\\ K. Deiters \etal, Nucl. Instr. Meth. {\bf A
  323} (1992) 162;\\ M. Chemarin \etal, Nucl. Instr. Meth. {\bf A 349} (1994)
  345;\\ M. Acciarri \etal, Nucl. Instr. Meth. {\bf A 351} (1994) 300;\\ G.
  Basti \etal, Nucl. Instr. Meth. {\bf A 374} (1996) 293;\\ A. Adam \etal,
  Nucl. Instr. Meth. {\bf A 383} (1996) 342;\\ O. Adriani \etal, Phys. Rep.
  {\bf 236} (1993) 1\relax
\relax
\bibitem{hzha}
HZHA version 2 is used. P.~Janot,
\newblock  in {\em Physics at LEP2}, ed. {{G.~Altarelli, T.~Sj\"ostrand
  and~F.~Zwirner}},  (CERN 96-01, 1996), volume~2, p. 309\relax
\relax
\bibitem{jetset73}
T. Sj{\"o}strand, CERN-TH 7112/93, CERN (1993), revised August 1995; \newline
  T. Sj{\"o}strand, Comp.\ Phys.\ Comm.\ {\bf 82} (1994) 74\relax
\relax
\bibitem{KORALW}
M.\ Skrzypek \etal, Comp. Phys. Comm. {\bf 94} (1996) 216;\\ M.\ Skrzypek
  \etal, Phys. Lett. {\bf B 372} (1996) 289\relax
\relax
\bibitem{PHOJET}
R.\ Engel, Z.\ Phys.\ {\bf C 66} (1995) 203; \newline R.\ Engel and J.\ Ranft,
  Phys.\ Rev.\ {\bf D 54} (1996) 4244\relax
\relax
\bibitem{DIAG36}
F.A.\ Berends, P.H.\ Daverfeldt and R.\ Kleiss, \NP {\bf B 253} (1985)
  441\relax
\relax
\bibitem{KORALZ40}
S. Jadach, B.F.L. Ward and Z. W\c{a}s, Comp. Phys. Comm. {\bf 79} (1994)
  503\relax
\relax
\bibitem{BHWIDE}
S.~Jadach \etal,
\newblock  Phys. Lett. {\bf B 390}  (1997) 298\relax
\relax
\bibitem{my_geant}
R. Brun \etal, {\em GEANT 3}, CERN DD/EE/84-1 (Revised), September 1987.\\ The
  GHEISHA program (H. Fesefeldt, RWTH Aachen Report PITHA 85/02, 1985) is used
  to simulate hadronic interactions\relax
\relax
\bibitem{nnet}
L3 Collab., M.~Acciarri \etal, Preprint CERN-EP/2000-104\relax
\relax
\bibitem{lnq}
ALEPH, DELPHI, L3 and OPAL Collab., The LEP Working Group for Higgs Boson
  Searches, Preprint CERN-EP/2000-055\relax
\relax
\bibitem{l3cl}
L3 Collab., O.~Adriani \etal, \PL {\bf B 411} (1997) 373\relax
\relax
\end{mcbibliography}
\end{document}